\documentclass[twocolumn,twocolappendix,dvipsnames]{aastex63}

\usepackage{amsmath}
\usepackage{comment}
\usepackage{xcolor}
\definecolor{amethyst}{rgb}{0.6, 0.4, 0.8}

\usepackage{lineno}

\received{June 1, 2019}
\revised{January 10, 2019}
\accepted{\today}

\shorttitle{Standalone VFISV}
\shortauthors{Gri\~n\'on-Mar\'in et al.}

\graphicspath{{./}{figures/}}

\begin{document}

\title{Improvement of the Helioseismic and Magnetic Imager (HMI) Vector Magnetic Field Inversion Code}

\correspondingauthor{A. B. Gri\~n\'on-Mar\'in}
\email{abgm@sun.stanford.edu / a.b.g.marin@uio.no}

\author[0000-0002-0570-4029]{Ana Bel\'en Gri\~n\'on-Mar\'in}
\affiliation{W. W. Hansen Experimental Physics Laboratory, Stanford University, Stanford, CA 94305-4085, USA}
\affiliation{Institute of Theoretical Astrophysics, University of Oslo, P.O. Box 1029 Blindern, NO-0315 Oslo, Norway}
\affiliation{Rosseland Centre for Solar Physics, University of Oslo, P.O. Box 1029 Blindern, NO-0315 Oslo, Norway}

\author[0000-0003-2359-9039]{Adur Pastor Yabar}
\affiliation{Institute for Solar Physics, Department of Astronomy,  Stockholm University, Albanova University Centre, SE-106 91 Stockholm, Sweden}

\author[0000-0002-0671-689X]{Yang Liu}
\affiliation{W. W. Hansen Experimental Physics Laboratory, Stanford University, Stanford, CA 94305-4085, USA}

\author[0000-0001-9130-7312]{J. Todd Hoeksema}
\affiliation{W. W. Hansen Experimental Physics Laboratory, Stanford University, Stanford, CA 94305-4085, USA}

\author[0000-0003-2622-7310]{Aimee Norton}
\affiliation{W. W. Hansen Experimental Physics Laboratory, Stanford University, Stanford, CA 94305-4085, USA}

\begin{abstract}

A spectral line inversion code, Very Fast Inversion of the Stokes Vector (VFISV), has been used since May 2010 to infer the solar atmospheric parameters from the spectropolarimetric observations taken by the Helioseismic and Magnetic Imager (HMI) aboard the Solar Dynamics Observatory (SDO). The magnetic filling factor, the fraction of the surface with a resolution element occupied by magnetic field, is set to have a constant value of one in the current version of VFISV. This report describes an improved inversion strategy for the spectropolarimetric data observed with HMI for magnetic field strengths of intermediate values in areas spatially not fully resolved. The VFISV inversion code has been modified to enable inversion of the Stokes profiles with two different components: one magnetic and one non-magnetic. In this scheme, both components share the atmospheric components except for the magnetic field strength, inclination, and azimuth. In order to determine whether the new strategy is useful, we evaluate the inferred parameters inverted with one magnetic component (the original version of the HMI inversion) and with two components (the improved version) using a Bayesian analysis. In pixels with intermediate magnetic field strengths (e.g. plages), the new version provides statistically significant values of filling fraction and magnetic field vector. Not only does the fitting of the Stokes profile improve, but the inference of the magnetic parameters and line-of-sight velocity are obtained uniquely. The new strategy is also proven to be effective for mitigating the anomalous hemispheric bias in the east-west magnetic field component in moderate field regions. 

\end{abstract}

\keywords{Sun: Instrumentation and Data Management; Sun: Magnetic fields, Photosphere}


\section{Introduction} 
\label{sec:intro}

A variety of magnetic structures can be observed at the solar surface, from the most impressive ones such as sunspots and filaments, to the smallest ones such as individual elements of the network or faculae. Studying the light coming from these features, we can ascertain some of their physical properties. Arguably, spectropolarimetry is the best tool to infer the magnetic, dynamic, and thermodynamic properties of the solar atmosphere where the observed spectral lines are formed. Since the early 1970s various approaches have been used to solve the radiative transfer equation to find a model that closely matches an observed quantity by optimizing a merit function. Solving the radiative transfer equation for polarized light poses strong theoretical and technical challenges leading to different degrees of approximations depending on the studied case \cite[see][for deeper insight into this topic]{deltoroiniesta2016}.

One of the simplest approximations for the solution of the radiative transfer equation that keeps a rigorous enough treatment of the magnetic field in the Zeeman regime is the Milne-Eddington (ME) approach \citep{landolfi1982}. The ME model assumes that the physical parameters in the atmosphere are independent of the optical depth except for the source function, which is assumed to vary linearly. While this is an oversimplified model, it has been widely used and leads to reasonable results in regions where field gradients are not too great.

Using a simple model generally imposes some limitations. For example, in the ME approximation the substructure of the resolution element is not treated, and that may influence the inferred atmospheric parameters. As shown by \cite{solanki1993}, there are different scenarios in which a single-component magnetic field model may be unable to properly model observed features: 1.- network or plage where magnetic and non-magnetic areas are mixed in the same resolution element; 2.- penumbral filaments where two magnetic structures with different physical properties coexist in the same pixel or areas where the magnetic field depends on height. In general, these scenarios are modeled using more than one ME component for each resolution element weighted by a factor calling filling factor. A better solution for these scenarios would be given by: 1.- two-component model with one field-free component; or 2.- two magnetic components.

This work focuses on the first case: a two-component model with one non-magnetic atmospheric component, as we aim to deal with spatially unresolved magnetic features outside of active regions. For this model, there are different scenarios depending on the magnetic field feature (weak, intermediate and strong) and the size of the structure compared to the spatial resolution \citep[for a more detailed discussion see][]{solanki1993}. An example of this can be found in \cite{harvey1972} who solves some discrepancies found during the interpretation of magnetic structures outside of sunspots by considering unresolved magnetic elements. They used Stokes I and V parameters observed in the visible spectral range near disk center together with a simplified two-component model (one magnetic and one non-magnetic weighted by a factor) to obtain the atmospheric parameters assuming a ME atmosphere during the inversion process. They concluded that, under some circumstances, it is unrealistic to treat the unresolved magnetic region as a single magnetic field vector occupying the whole pixel. 

The filling factor provides rich information about the magnetic field, and fitting it in an inversion affects the resultant field strength, field orientation, and flux distribution, i.e. the entire magnetic configuration. It has also been suggested recently that lack of determination of filling factor leads to a mismatch in the line-of-sight and transverse magnetic field components inferred by inversion processing, and that this ultimately causes the hemisphere bias found in the east-west field component in the moderate field strength regions \citep{liu2021}. This bias, seen in the full-disk vector magnetograms from both the Helioseismic and Magnetic Imager (HMI; \citealt{scherrer2012}, \citealt{schou2012}) on board the Solar Dynamics Observatory (SDO; \citealt{pesnell2012}) satellite and the Vector Stokes Magnetograph (VSM) on the ground-based Synoptic Optical Long-term Investigation of the Sun \cite[SOLIS;][]{keller2003}, has been described and discussed in \citet{pevtsov2021}. It is suggested that this systematic bias may be caused by a combination of the noise in measurements and the magnetic filling factor that is not resolved accurately. Thus it is timely to improve the Very Fast Inversion of the Stokes Vector (VFISV, \citet{borrero2011b,centeno2014}) code used to invert HMI data, so that it can properly determine the filling factor.

In this work we investigate a more complex model for magnetic features outside of active regions that is based on VFISV and can determine filling factor from HMI observations. The goals of this study are (1) to improve the inversion approach for HMI data to derive more accurate magnetic field data products generally, and (2) to mitigate the systematic hemispheric bias in the east-west component of magnetic field in moderate-field regions.

The paper is organized as follows. Section 2 describes the old and new models used to infer the atmospheric parameters with VFISV inversion code. In section 3, we evaluate the inferred parameters with both models using a Bayesian analysis. In section 4, we show examples of mitigation of the hemisphere bias in moderate-field regions in vector magnetic field inverted using the new version of VFISV. And finally, in section 5, we discuss the results of our work and show the conclusions.


\section{Methods and/or Model} 
\label{sec:model}

\noindent

HMI/SDO is a filtergraph with full disk coverage at 4096$\times$4096 pixels and it continuously records the evolution of the visible solar surface. In particular, HMI observes the full Stokes parameters at six different wavelengths around the magnetically sensitive Fe\,{\sc i} 6173 {\AA} spectral line with a spectral resolution of around $R=85000$ and a spatial sampling of 0\farcs504 \citep{schou2012}. It takes 135 seconds (90 seconds after April 2016 in mod-L mode) to obtain a set of filtergrams in 6 polarization states at 6 wavelength positions. The Stokes parameters [I, Q, U, V] are computed from those measurements. In order to suppress the p modes and increase the signal-to-noise ratio, usually the Stokes parameters are derived from the filtergrams averaged over certain time. Currently the average is computed every 720-seconds \citep{Hoeksema2014}. The Stokes parameters are then inverted assuming an ME model with a uniform one-component magnetic field and filling factor fixed to 1. The VFISV inversion solves for the vector magnetic field (magnetic field strength, B, inclination, $\gamma$, and azimuth, $\Psi$
\footnote{The inclination is measured with respect to the line-of-sight, and the reference for $0^\circ$ azimuth is in the north-south solar direction, and the azimuth values increase in a counterclockwise direction})
and other model parameters, including the absorption coefficient, $\eta_{0}$, the damping, $a$, the Doppler width, $\Delta\lambda_{D}$, the source function at the base of the photosphere, $\rm S_{0}$, the gradient of the source function, $\rm S_{1}$, and the line-of-sight velocity, $\rm V_{\rm los}$:

\begin{equation}
\label{eq:models}
    {\rm\mathbf I}^{\rm syn}_{\rm 1c} = {\rm\mathbf I}_{\rm m}^{\rm syn}(\eta_{0}, \gamma, \Psi, \Delta\lambda_{D}, B, {\rm V}_{\rm los}, {\rm S}_{0}, {\rm S}_{1})
\end{equation}

\noindent
In the current version of VFISV, The damping parameter is fixed to 0.5. This way, the code has 8 free parameters.

In order to better characterize magnetic structures embedded in a magnetic-free atmosphere, we now include a second non-magnetic component. After considering different strategies, we conclude that the best performance is achieved when the non-magnetic component is modeled using the same non-magnetic model parameters ($\eta_{0}$, $a$, $\Delta\lambda_{D}$, ${\rm V}_{\rm los}$, $\rm S_{0}$, $\rm S_{1}$) as the magnetic model, and we weight its relative contribution by means of a geometrical parameter (1 - $\alpha_{\rm mag}$) that defines the magnetic fraction of the resolution element.

This two-component model is still highly simplified, as only one additional free parameter is added (the model has 9 free parameters), the magnetic filling factor ($\alpha_{\rm mag}$). We use this approach in order to be able to consider a two component approach while at the same time, keeping it as simple as possible given that the HMI Stokes profiles provide limited information (i.e. we have access to 24 values, only six wavelengths per Stokes parameter). In mathematical formulation, our two-component model is:

\begin{equation}\label{eq:model2c}
    {\rm\mathbf I}^{\rm syn}_{\rm 2c}=\alpha_{\rm mag} {\rm\mathbf I}_{\rm m}^{\rm syn} + (1 - \alpha_{\rm mag}) {\rm\mathbf I}_{\rm nm}^{\rm syn},
\end{equation}

\noindent
where both ME components depend on the same non-magnetic parameters:

\begin{align}
    {\rm\mathbf I}^{\rm syn}_{\rm m} &= \mathcal{M}(\eta_{0}, \gamma, \Psi, \Delta\lambda_{D}, B, \rm V_{\rm los}, \rm S_{0}, \rm S_{1})\\
    {\rm\mathbf I}^{\rm syn}_{\rm nm} &= \mathcal{M}(\eta_{0}, \Delta\lambda_{D}, \rm V_{\rm los}, \rm S_{0}, \rm S_{1})
\end{align}    

The modification to the model requires that the derivatives of the new model observables (synthetic Stokes profiles) be computed with respect to the model parameters. Now, the new derivatives are calculated as an algebraic operation on the pre-existing response functions. This is because VFISV looks for the set of model parameters that best fit the observations using a Levenberg–Marquardt \cite[LM;][]{levenberg1944,marquadt1963} least-squares fitting algorithm. Starting from an initial guess, the procedure iteratively modifies the model output to determine the set of parameters that best fit the observations. To do so, the LM-algorithm makes use of these derivatives, also known as the response function in the context of the radiative transfer equation. According to the modification of the ME-model (see Equation \ref{eq:model2c}) the response functions are changed as follows:
\begin{align}
    \frac{\partial {\rm\mathbf I}^{\rm syn}_{\rm 2c}}{\partial \mathcal{M}_{\rm i}}&=
    \frac{\partial \alpha_{\rm mag} {\rm\mathbf I}_{\rm m}^{\rm syn}}{\partial \mathcal{M}_{\rm i}} + 
    \frac{\partial (1 - \alpha_{\rm mag}) {\rm\mathbf I}_{\rm nm}^{\rm syn}}{\partial \mathcal{M}_{\rm i}}\\
    &\!\!\!\!\!\!\left\{
            \begin{array}{ll}
                      ={\rm\mathbf I}_{\rm m}^{\rm syn} - \alpha_{\rm mag} {\rm\mathbf I}_{\rm nm}^{\rm syn} \\
                      \qquad\qquad if:\ \mathcal{M}_{\rm i}=\alpha_{\rm mag}\\
                      \\
                      =\alpha_{\rm mag} \frac{\partial {\rm\mathbf I}_{\rm m}^{\rm syn}}{\partial \mathcal{M}_{\rm i}}\\
                      \qquad\qquad if:\ \mathcal{M}_{\rm i}=B,\gamma,\Psi\\
                      \\
                      =\alpha_{\rm mag} \frac{\partial {\rm\mathbf I}_{\rm m}^{\rm syn}}{\partial \mathcal{M}_{\rm i}} + (1 - \alpha_{\rm mag}) \frac{\partial {\rm\mathbf I}_{\rm nm}^{\rm syn}}{\partial \mathcal{M}_{\rm i}}\\
                      \qquad\qquad if:\ \mathcal{M}_{\rm i}={\rm S}_{\rm 0}, {\rm S}_{\rm 1}, \eta_{\rm 0}, \Delta\lambda_{\rm D}, {\rm V}_{\rm los} \\
                \end{array}
              \right.
\end{align}


\section{Reliability assessment}
\label{sec:mcmc}

The increase in the model complexity of ${\rm\mathbf I}_{\rm m}^{\rm syn}$ might imply strong degeneracies among model parameters. It is also important to recognize that the two-component model (${\rm\mathbf I}^{\rm syn}_{\rm 2c}$) simply fits to more parameters than the one-component model (${\rm\mathbf I}^{\rm syn}_{\rm 1c}$), so in order to properly ascertain which model better fits the observations, we need to account for the difference in the number of degrees of freedom. Among the various approaches that might be taken to quantify the performance, we have chosen to use a Bayesian approach. This approach allows for full characterization of the posterior distribution of the various model parameters (and uncertainties) as well as their possible degeneracies, and it also allows a rigorous comparison of models with different number of free parameters. This report does not discuss the specifics of Bayesian inference, we simply summarize the most relevant features for our situation and refer the interested reader to more complete review \citep[for instance][]{gregory2005}.

According to Bayes theorem \citep{bayes1763}, provided a model (${\mathbf I}_{j}^{\rm syn}$) and an observation (${\mathbf I}^{\rm obs}$) the posterior distribution of the model parameters is given by:

\begin{equation}\label{eq:bayes}
    p(\mathcal{M}_{\rm i}|{\mathbf I}^{\rm obs},{\mathbf I}_{j}^{\rm syn})=
    \frac
    {p({\mathbf I}^{\rm obs}|\mathcal{M}_{\rm i},{\mathbf I}_{j}^{\rm syn})
    \,p(\mathcal{M}_{\rm i}|{\mathbf I}_{j}^{\rm syn})}
    {\int p({\mathbf I}^{\rm obs}|\mathcal{M}_{\rm i},{\mathbf I}_{j}^{\rm syn})
    \,p(\mathcal{M}_{\rm i}|{\mathbf I}_{j}^{\rm syn})\,d\mathcal{M}_{\rm i}},
\end{equation}

\noindent
where $p({\mathbf I}^{\rm obs}|\mathcal{M}_{\rm i},{\mathbf I}_{j}^{\rm syn})$ stands for the likelihood likelihood of observations ${\mathbf I}^{\rm obs}$ conditional on the model and synthesis ${\mathbf I}^{\rm syn}$, and $p(\mathcal{M}_{\rm i}|{\mathbf I}_{j}^{\rm syn})$ is the information about the model $\mathcal{M}_{\rm i}$ parameters that is supplied, i.e. the prior. The denominator is known as the evidence (\mbox{$E_{j}=\int p({\mathbf I}^{\rm obs}|\mathcal{M}_{\rm i},{\mathbf I}_{j}^{\rm syn})\,p(\mathcal{M}_{\rm i}|{\mathbf I}_{j}^{\rm syn})\,d\mathcal{M}_{\rm i}$}) and, while it can be omitted when using a single model, it is key to making a proper comparison between two different models trying to explain the same observation, as it takes into account the size of the hyper-space of the model parameters. Here, sub-index $j$ is used to make explicit that we will use this approach for both models ($\mathbf{\rm I}_{\rm 2c}^{\rm syn}$) and ($\mathbf{\rm I}_{\rm 1c}^{\rm syn}$).

In the Bayesian approach the aim is to determine the posterior distributions for the model parameters, so some likelihood functional and a prior must be provided. Regarding the prior, we set an informative prior in which all but one of the model parameters $\mathcal{M}$ are equally probable in a certain range of the parameter space while impossible outside it. For $\eta_{0}$ the prior is normally distributed. The specific values for each parameter prior are gathered in Table \ref{tab:table_prior_ranges}. This configuration is chosen following the actual implementation of VFISV.

\begin{deluxetable}{llll}[t]
\caption{Model parameter priors employed for the Bayesian inference. For uniform priors minimum and maximum values are provided. For $\eta_{0}$ the prior is a normal distribution around a mean value of five and a standard deviation of $\sqrt{1/2}$. $\alpha_{\rm mag}^\dagger$ (filling factor) is only used for the two-component model.}
\label{tab:table_prior_ranges}
\tablehead{$\mathcal{M}_{\rm i}$ & $p(\mathcal{M}_{\rm i}|{\mathbf I}_{j}^{\rm syn})$ & Units} 
\startdata
    B & $\mathcal{U}(5, 5000)$ & G \\
    $\gamma$ & $\mathcal{U}(0, 180)$ & deg \\
    $\Psi$ & $\mathcal{U}(0, 180)$ & deg \\
    $\Delta\lambda_{\rm D}$ & $\mathcal{U}(1, 500)$ & m{\AA}\\
    $\eta_{\rm 0}$ & $\Gamma(5,\sqrt{1/2})$ & - \\
    ${\rm V}_{\rm los}$ & $\mathcal{U}(-7, 7)$ & km/s\\
    ${\rm S}_{0}$ & $\mathcal{U}(0.15, 1.2)$ & $\mathbf{I}_{1,{\rm cont}}^{\rm obs}$ \\
    ${\rm S}_{1}$ & $\mathcal{U}(0.15, 1.2)$ & $\mathbf{I}_{1,{\rm cont}}^{\rm obs}$ \\
    $\alpha_{\rm mag}^{\dagger}$ & $\mathcal{U}(0, 1)$ & -
\enddata
\end{deluxetable}

\begin{figure*}[!ht]
  \includegraphics[width=\linewidth]{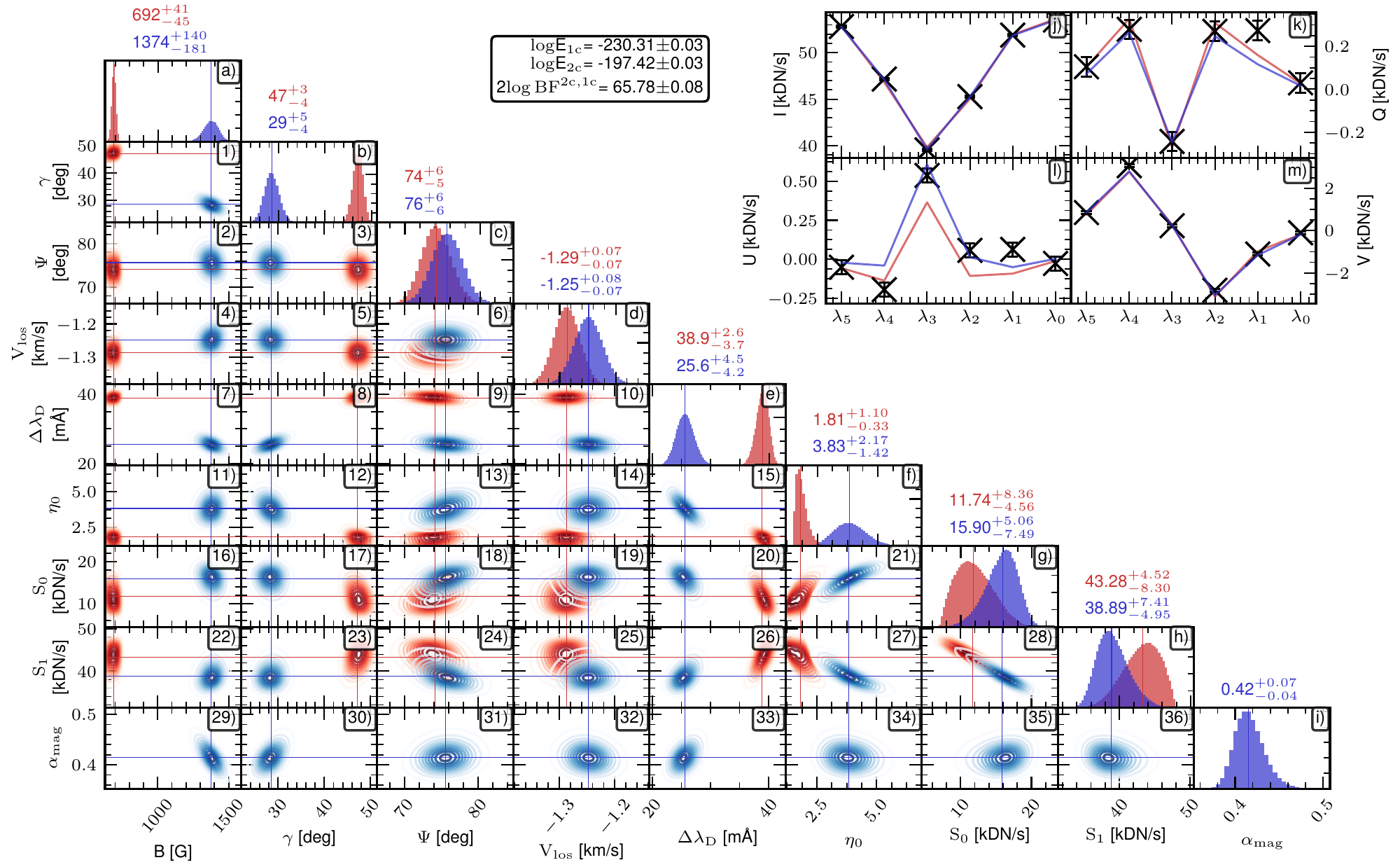}
  \caption{Bayesian inference for the single-component (red) and two-component (blue) models for a plage pixel at heliocentric angle $\theta=55.25$ deg. Panels from {\it (1)} to {\it (36)} show the two-dimensional probability density functions for the model parameter combinations. Panels {\it (a)} to {\it (i)} on the diagonal show the marginal posterior distributions for each model parameter independently. The color coded values show the median value (color coded vertical thin lines) for each model as well as their 99.73\% credible region. Panels {\it (j)} to {\it (m)} at the upper right depict the measured Stokes spectra (black crosses) with their uncertainties (vertical black bars) together with the synthetic profiles for the maximum-likelihood case for the single-component model (red; ${\rm S}_{0}=6.91$ kDN/s, ${\rm S}_{1}=48.0$ kDN/s, $\eta_{0}=1.48$, $\Delta\lambda_{\rm D}=40.0$ m{\AA}, $B=695$ G, $\gamma=47$ deg, $\Psi=73$ deg, and ${\rm V}_{\rm los}=-1.27$ km/s) and two-component model (blue; ${\rm S}_{0}=8.92$ kDN/s, ${\rm S}_{1}=45.77$ kDN/s, $\eta_{0}=2.44$, $\Delta\lambda_{\rm D}=28.20$ m{\AA}, $B=1383$ G, $\gamma=29$ deg, $\Psi=72$ deg, ${\rm V}_{\rm los}=-1.25$ km/s, and $\alpha_{\rm mag}=0.41$). The text box shows the actual evidences found (in natural logarithm scale) together with the Bayes factor.}
  \label{fig:plage_c4_case}
\end{figure*}

The likelihood function is set to be uncorrelated and of Gaussian form with zero mean and an observing uncertainty of $\sigma_{2,3}=47$ kDN/s, as derived from full-disk close-to-continuum polarimetric data. In addition, Stokes I and V are allowed worse fitting by means of weights, so that they have an effective uncertainty of $\sigma_{1}=120$ and $\sigma_{4}=69$ kDN/s, respectively. These weights are chosen in order to mimic the VFISV weightings used in the standard pipeline processing. In mathematical form, this likelihood is expressed as:

\begin{equation}
    p({\mathbf I}^{\rm obs}|\mathcal{M}_{\rm i},{\mathbf I}_{j}^{\rm syn})=
    \prod_{s,l}\left\{
    \frac{1}{\sqrt{2\,\pi\,\sigma_{\rm s}^{2}}}
    \exp{
    \left[-\frac{({\mathbf I}^{\rm obs}_{s,l}-{\mathbf I}^{\rm syn}_{j;s,l})^{2}}{2\,\sigma^{2}_{s}}\right]
    }\right\},
\end{equation}

\noindent
where sub-index $s$ is used for the four Stokes parameters and sub-index $l$ for the six wavelength tuning positions of HMI. The uncertainty for each Stokes parameter is the same for all the tuning positions.

Provided with functional forms for the likelihood and prior, the estimation of the posterior distribution of the model parameters ($\mathcal{M}$) as well as the evidence (${\rm E}_{j}$) is done using PyMultiNest \citep{buchner2014}, which internally uses the Multinest library \citep{feroz2008,feroz2009,feroz2019}. The sampling process is extremely computationally demanding, so we have used it over a handful of representative cases in order to evaluate the model parameter inference ability of the two-component model, some of which are shown below.

The comparison of two models is essentially an extension of the formalism detailed in \cite{kass1995,arregui2015}. Accordingly, the two models can be compared by using the Bayes factor (${\rm BF}$) which, for the two models considered here, is given by: 

\begin{equation}
    {\rm BF}^{2c,1c}=\frac
    {p({\mathbf I}^{\rm obs}|{\mathbf I}^{\rm syn}_{2c})\,p({\mathbf I}^{\rm syn}_{2c})}
    {p({\mathbf I}^{\rm obs}|{\mathbf I}^{\rm syn}_{1c})\,p({\mathbf I}^{\rm syn}_{1c})},
\end{equation}

\noindent
where $p({\mathbf I}^{\rm obs}|{\mathbf I}^{\rm syn}_{j})$ stands for the likelihood of model $j$ provided the observable ${\mathbf I}^{\rm obs}$ (notice that this is the evidence as it appears in Equation\,(\ref{eq:bayes}) and $p({\mathbf I}^{\rm syn}_{j})$ is the prior for that model. Here, we assume that both models have the same {\it a priori} probability, so that they cancel each other, and the model comparison ends up comparing the evidence estimated for each model. It is important to notice that this is a probabilistic comparison, so rather than having a threshold value in order to state which model explains the observation better, some degree of evidence is defined depending on the value of ${\rm BF}$. Here, we follow \cite{kass1995} in using a natural logarithmic table, which is presented in Table\,\ref{tab.:kass_evidence_levels} for the sake of completeness.

\begin{deluxetable}{cl}[t]
\caption{Values for Bayes factors favoring the two-component model vs. the single component one \citep[following][]{kass1995}.}
\label{tab.:kass_evidence_levels}
\tablehead{$2\,\log{{\rm BF}^{2c,1c}}$ & Evidence Favoring ${\mathbf I}^{\rm syn}_{2c}$} 
\startdata
    0,2 & Not significant\\
    2,6 & Positive \\
    6,10 & Strong\\
    $\ge$10 & Very strong
\enddata
\end{deluxetable}

\begin{figure*}
  \includegraphics[width=1\linewidth]{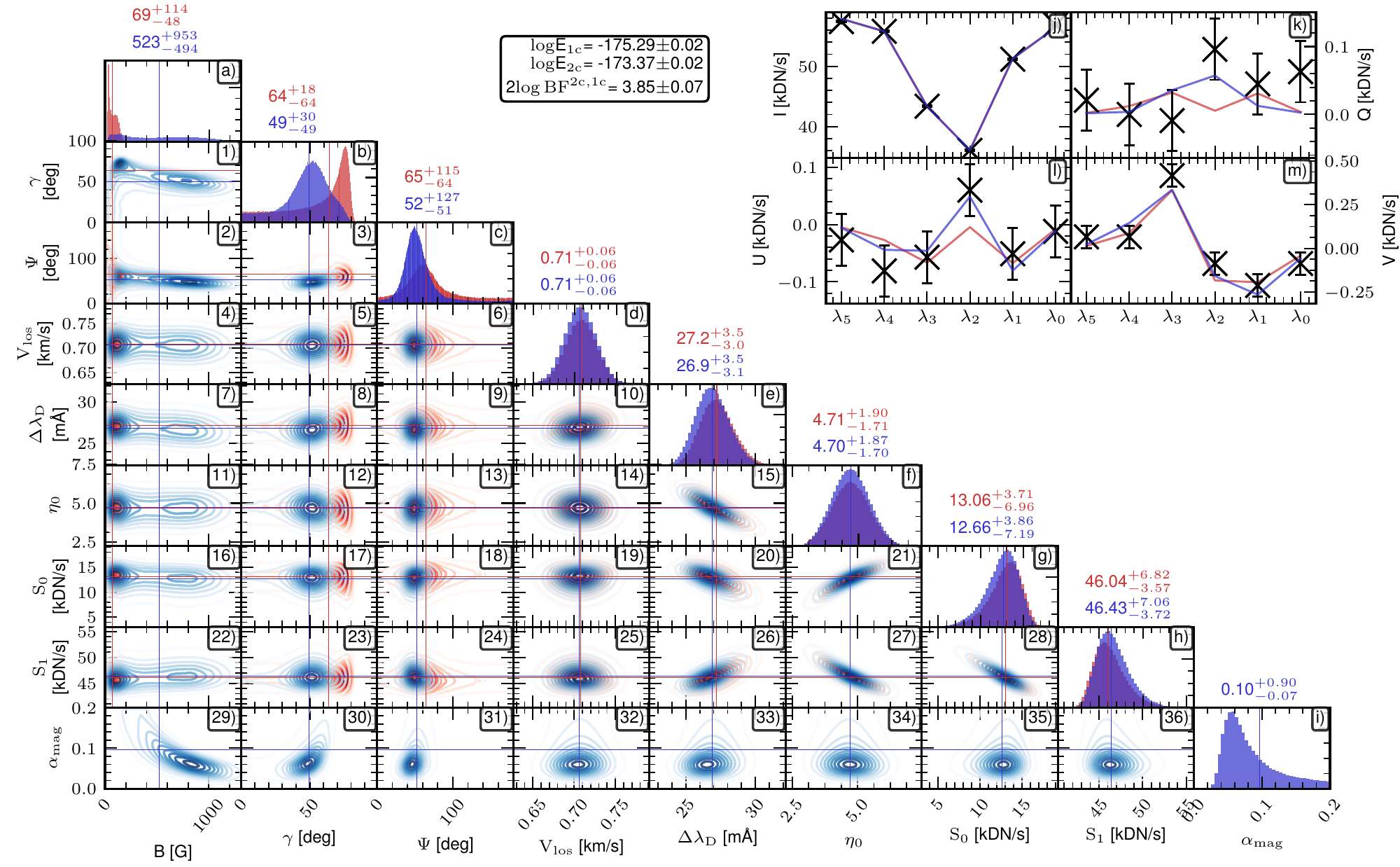}
  \caption{As in Figure\,\ref{fig:plage_c4_case} for a low-magnetized quiet-Sun pixel. The specific model parameter values for the fits shown in panels {\it (j)} to {\it (l)} are, for the single-component model (red; ${\rm S}_{0}=7.06$ kDN/s, ${\rm S}_{1}=51.93$ kDN/s, $\eta_{0}=3.16$, $\Delta\lambda_{\rm D}=29.72$ m{\AA}, $B=146$ G, $\gamma=79$ deg, $\Psi=57$ deg, and ${\rm V}_{\rm los}=0.70$ km/s) and for the two-component model (blue; ${\rm S}_{0}=5.72$ kDN/s, ${\rm S}_{1}=53.30$ kDN/s, $\eta_{0}=3.04$, $\Delta\lambda_{\rm D}=29.61$ m{\AA}, $B=962$ G, $\gamma=49$ deg, $\Psi=44$ deg, ${\rm V}_{\rm los}=0.71$ km/s, and $\alpha_{\rm mag}=0.58$).}
  \label{fig:ex2}
\end{figure*}

The two-component model presented here aims at being able to deal with cases in which a strong, spatially unresolved magnetic structure is embedded in a (nearly) field-free plasma. Clear candidates for this model are plage and network regions. Such regions host small-scale $\approx0^{\prime\prime}$.1-100km \citep{lagg2010} relatively strong magnetic fields where a model with two components, one magnetic ($\mathbf{\rm I}_{\rm m}^{\rm syn}$) and one field-free ($\mathbf{\rm I}_{\rm nm}^{\rm syn}$), may perform better.

An example is shown in Figure\,\ref{fig:plage_c4_case} in which we present the result of the Bayesian approach for a single plage pixel ($\mathbf{I}^{\rm obs}$). In the lower left panels, the one-dimensional and two-dimensional posterior distributions of the model parameters for the one-component model ($\mathbf{I}^{\rm syn}_{1c}$ in red) and the two-component model ($\mathbf{I}^{\rm syn}_{2c}$ in blue) are shown. From the marginal posterior distribution for each model (panels (a) to (j)) it is evident that both models are able to infer a unique parameter set (at least those related to the magnetic field vector), i.e. the various model-parameter one-dimensional marginal distributions are relatively narrow for most of the model parameters. The two-dimensional marginal distributions (panels (1) to (36)) show that some degeneracies exist between some of the model parameters. For instance, ${\rm S}_{0}$ and ${\rm S}_{1}$ are inverse linearly related, i.e. if ${\rm S}_{0}$ increases its value, ${\rm S}_{1}$ becomes proportionally smaller, keeping a compatible likelihood. Another example is the degeneracy between $\Delta\lambda_{\rm D}$ and $\eta_{0}$ that follows a more complex relation. These two examples are known to occur for the HMI spectropolarimetric observations \citep{centeno2014}. Some additional degeneracies seem to exist. For instance $\eta_{0}$ and ${\rm S}_{0}$ and ${\rm S}_{1}$ (panels (21) and (27)). In particular, as $\eta_{0}$ increases, so does ${\rm S}_{0}$, while ${\rm S}_{1}$ gets smaller. Another degeneracy is present between $\Delta\lambda_{\rm D}$ and ${\rm S}_{0}$ and ${\rm S}_{1}$ (panels (20) and (26)). Despite the presence of these degeneracies, the magnetic field parameters and line-of-sight velocity seem to be fairly decoupled ($\psi$ does show a small degeneracy with ${\rm S}_{0}$ and ${\rm S}_{1}$, but its value stays confined to $<12^\circ$ with 99.73\% certainty). To this point both models $\mathbf{I}^{\rm syn}_{1c}$ and $\mathbf{I}^{\rm syn}_{2c}$ are quite similar in their behaviors, though the actual values for the distributions vary. 

However, when considering $\mathbf{I}^{\rm syn}_{2c}$ the additional parameter ($\alpha_{\rm mag}$) (panels (29) to (36) and (i)) must be considered. It is interesting to see that there is still some degree of degeneracy among $B$, $\gamma$ and $\alpha_{\rm mag}$. This is related to the fact that when the magnetic field is not in the strong-field regime, the model is not sensitive to each parameter individually, but to $\phi=\alpha_{\rm mag}\,B\,\cos{\gamma}$ \citep[the line-of-sight component of the mean magnetic flux density, see, for instance, ][]{asensio2007}. Thus, there exists an interplay between $\alpha_{\rm mag}$, $B$, and $\gamma$ while keeping the same likelihood. However, this degeneracy remains relatively small, for instance, the magnetic field strength is within a 300 G range with 99.73\% certainty instead of extending to very low values of magnetic field strengths. This result is due to the fact that there is some amount of linear polarization that poses additional constraints in the inference even when the magnetic field strength is weak \citep[see, for instance, ][ for a similar case in which linear polarization signals allow inferring the magnetic field orientation]{pastoryabar2020}.

The discussion so far has focused on the model parameter inference, but what we really want to know is whether one of the parametric models is preferred. To do so we follow \cite{kass1995} as shown in Table\,\ref{tab.:kass_evidence_levels}. For this plage case, the natural logarithm of the evidence for $\mathbf{I}^{\rm syn}_{1c}$ is $-230.31\pm0.03$ and for $\mathbf{I}^{\rm syn}_{2c}$ it is $-197.42\pm0.03$. Thus, the evidence ($2\log{BF^{2c,1c}}=65.78\pm0.08$) strongly favors $\mathbf{I}^{\rm syn}_{2c}$ in the plage, i.e. the result for the two-component model is to be preferred. We also note that the orientation and strength of the magnetic field vector are different for the two models considered here.

For comparison, Figure \ref{fig:ex2} shows the same model comparison for a nearby quiet-Sun pixel. Similarly to the plage case, the non-magnetic model parameters other than ${\rm V}_{\rm los}$ (${\rm S}_{0}$, ${\rm S}_{1}$, $\eta_{0}$, $\Delta\lambda_{D}$) show degeneracies. For the quiet-Sun case, we also see clear degeneracies between $B$ and $\gamma$ for the $\mathbf{I}^{\rm syn}_{1c}$ case and between $B$, $\gamma$, and $\alpha_{\rm mag}$ for $\mathbf{I}^{\rm syn}_{2c}$ (see panels (1), (29) and (30)). This is the expected behavior in the weak field regime and in the absence of linear polarization signals, where the inferred parameter is the line-of-sight magnetic flux density $\phi={\rm B}\,\cos{\gamma}$ (or $\phi=\alpha_{\rm mag}\,{\rm B}\,\cos{\gamma}$ in the $\mathbf{I}^{\rm syn}_{2c}$ case). The azimuth parameter ($\Psi$) is hardly-defined as there is no significant linearly polarized signal. For the quiet-Sun case, we find only a slight preference for the $\mathbf{I}^{\rm syn}_{2c}$ as $2\,\log{\rm BF}^{2{\rm c},1{\rm c}}=3.85\pm0.07$. It is important to note here that this behavior is expected for any inversion case outside the strong-field regime without linear polarization signals. In particular, for plage regions close to disk center (i.e. with low or no polarization signals at all) there is no unique determination of B and $\gamma$ (and $\alpha_{\rm mag}$ for the inversion with filling factor as a free parameter).

\begin{figure}
  \includegraphics[width=1\linewidth]{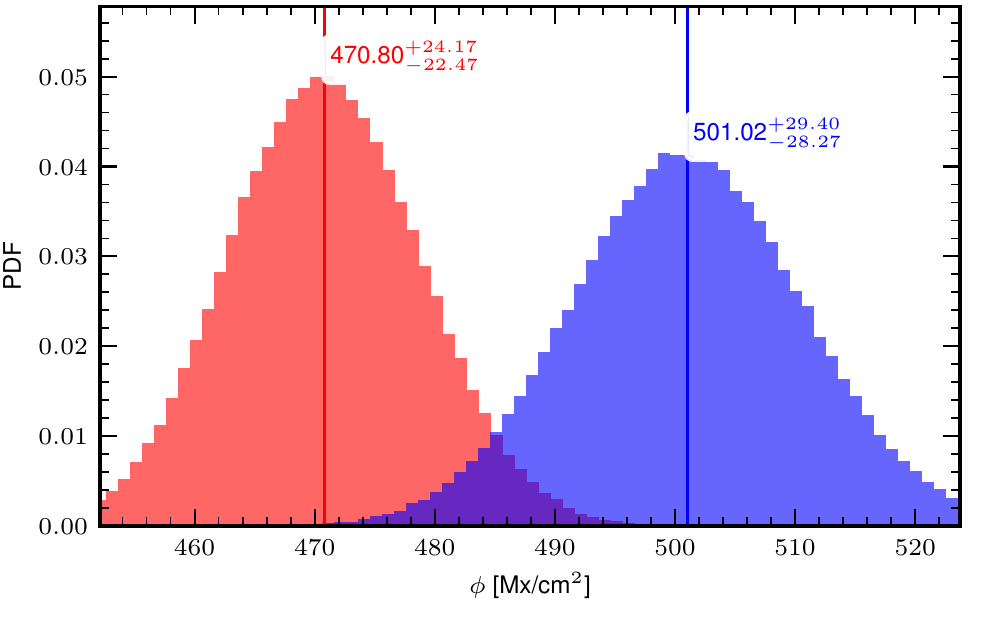}
  \includegraphics[width=1\linewidth]{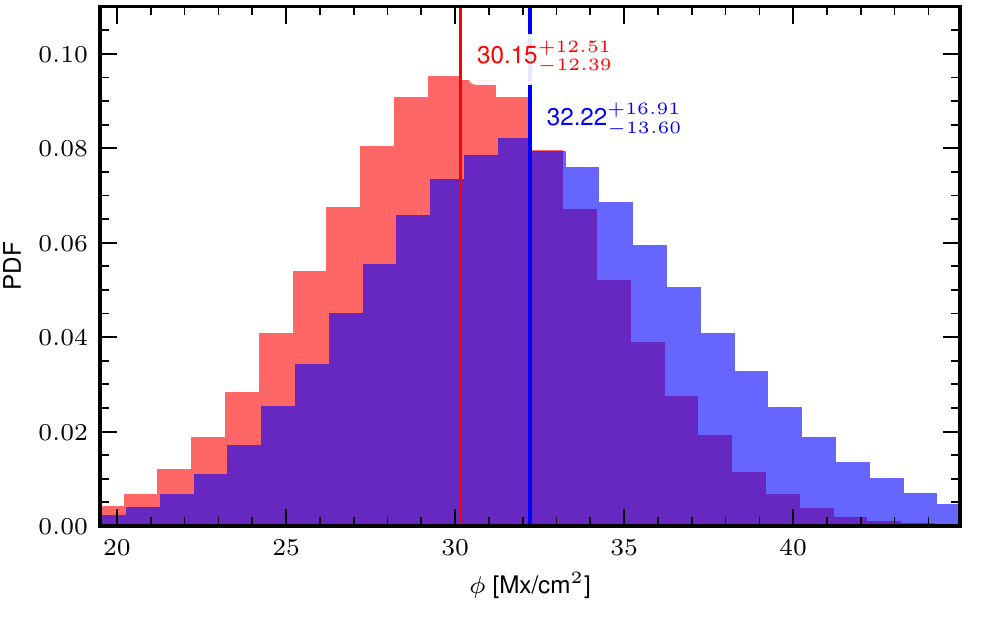}
  \caption{Probability density function for the inferred line-of-sight magnetic flux density ($\phi$) for the plage presented in Figure \ref{fig:plage_c4_case} (top panel) and for the quiet-Sun shown in Figure \ref{fig:ex2} (bottom panel). Results for the one-component ME model (red) and for the two-component model (blue) are shown. Sub- and super-indexes refer to the limits for the 99.97\% confidence interval.}
  \label{fig:fluxes}
\end{figure}

Figure \ref{fig:fluxes} shows the line-of-sight magnetic flux density calculated as $\phi=\alpha_{\rm mag}\,{\rm B}\,\cos{\gamma}$ for the selected pixels in Figure~\ref{fig:plage_c4_case} (plage, top panel) and Figure~\ref{fig:ex2} (quiet-Sun, bottom panel). The magnetic flux values obtained in quiet-Sun pixels are similar, whether the Stokes parameters are inverted with one component or with two. This is because the magnetic field strength and the inclination (and the filling factor too for the two-component model) are not inferred independently, i.e. the actual inferred parameter is the mean line-of-sight magnetic flux density. In plage pixels the magnetic flux density inferred using the one-component and two-component models are different, i.e. the two-component inversion leads to a different magnetic field orientation and greater line-of-sight magnetic flux density. The value is $\approx 30\,Mx/cm^{2}$) smaller when the pixel is inverted with only one component (the median is $\approx 470\,Mx/cm^{2}$ compared to $\approx 500\,Mx/cm^{2}$).

\section{Mitigation of the hemisphere bias in medium and weak magnetic field}

\begin{figure*}[ht]
\centering
  \includegraphics[width=0.9\linewidth]{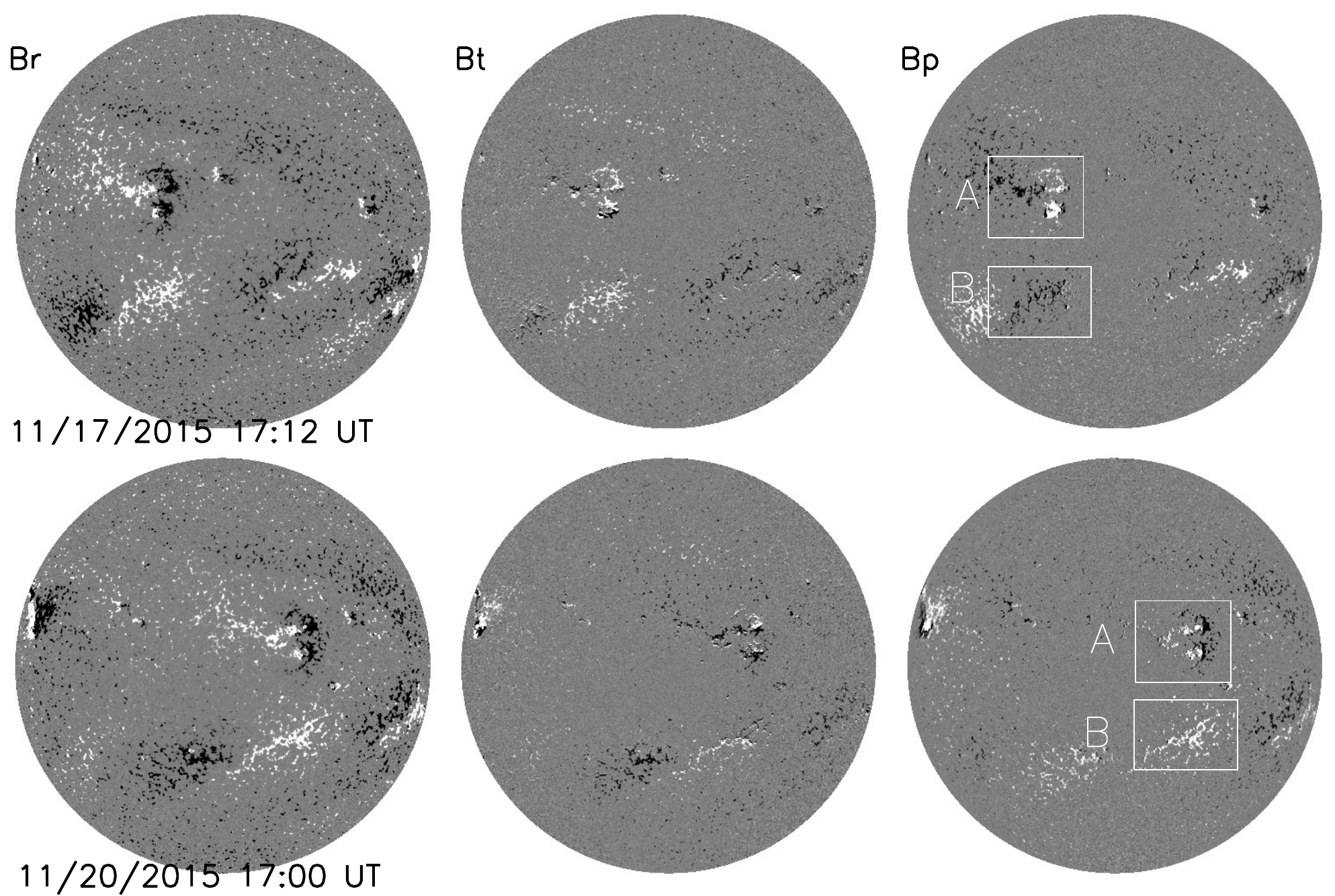}
  \caption{Full-disk HMI vector magnetograms taken at 17:12 UT on 17 November (top) and 17:00 UT on 20 November, 2015. The color table saturates at $\pm 60$ Mx cm$^{-2}$. Images from left to right show three components of magnetic field, $B_r$ (left; radial component), $B_t$ (middle; north-south component), and $B_p$ (right; east-west component). The vector field data are derived using the original version of VFISV. ``A'' and ``B'' denote two medium-strength field regions where the east-west field, $B_p$, changes sign after crossing central meridian.}
  \label{fig:fullDisk}
\end{figure*}

There is a hemispheric bias in medium-strength field regions seen in vector magnetograms from both VSM/SOLIS and HMI/SDO as reported by \citet{pevtsov2021}. The bias refers to a change in the sign of the heliographic east-west component of the magnetic field when a region crosses central meridian. Figure \ref{fig:fullDisk} shows an example. The top panels show the three components of magnetic field on the solar disk, $B_r$ (radial component), $B_t$ (north-south component), and $B_p$ (east-west component), observed at 17:12 UT, November 17, 2015. The bottom panels are the vector components measured at 17:00 UT, November 20, 2015, about three days later. The data are derived by applying VFISV to Stokes [I, Q, U, V] with a constant filling factor ($\alpha_{\rm mag} = 1$). The 180$^\circ$ ambiguity of the azimuth in the inverted magnetic field is resolved using the ``minimum energy'' algorithm \citep{metcalf1994,leka2009}. Two fairly stable regions, A and B, are identified in the $B_p$ panels on the right. Looking first at the radial-component panels on the left, Region A in the north has a negative-polarity feature toward the right with a less concentrated medium-strength positive polarity pattern toward the left. Region B in the south is entirely medium-strength and has a positive radial-field component throughout.
Now, consider the panels that show $B_p$. The east-west component of the southern medium-strength magnetic field in Region B changes sign from negative (black, upper-right panel) to positive (white, lower-right panel) after it rotates past central meridian.
Similarly, the medium-strength portion of Region A in the northern hemisphere changes sign from negative to positive. The sign of the most leading negative feature changes from positive to negative, as well.

\begin{figure*}[ht]
\centering
  \includegraphics[width=0.7\textwidth]{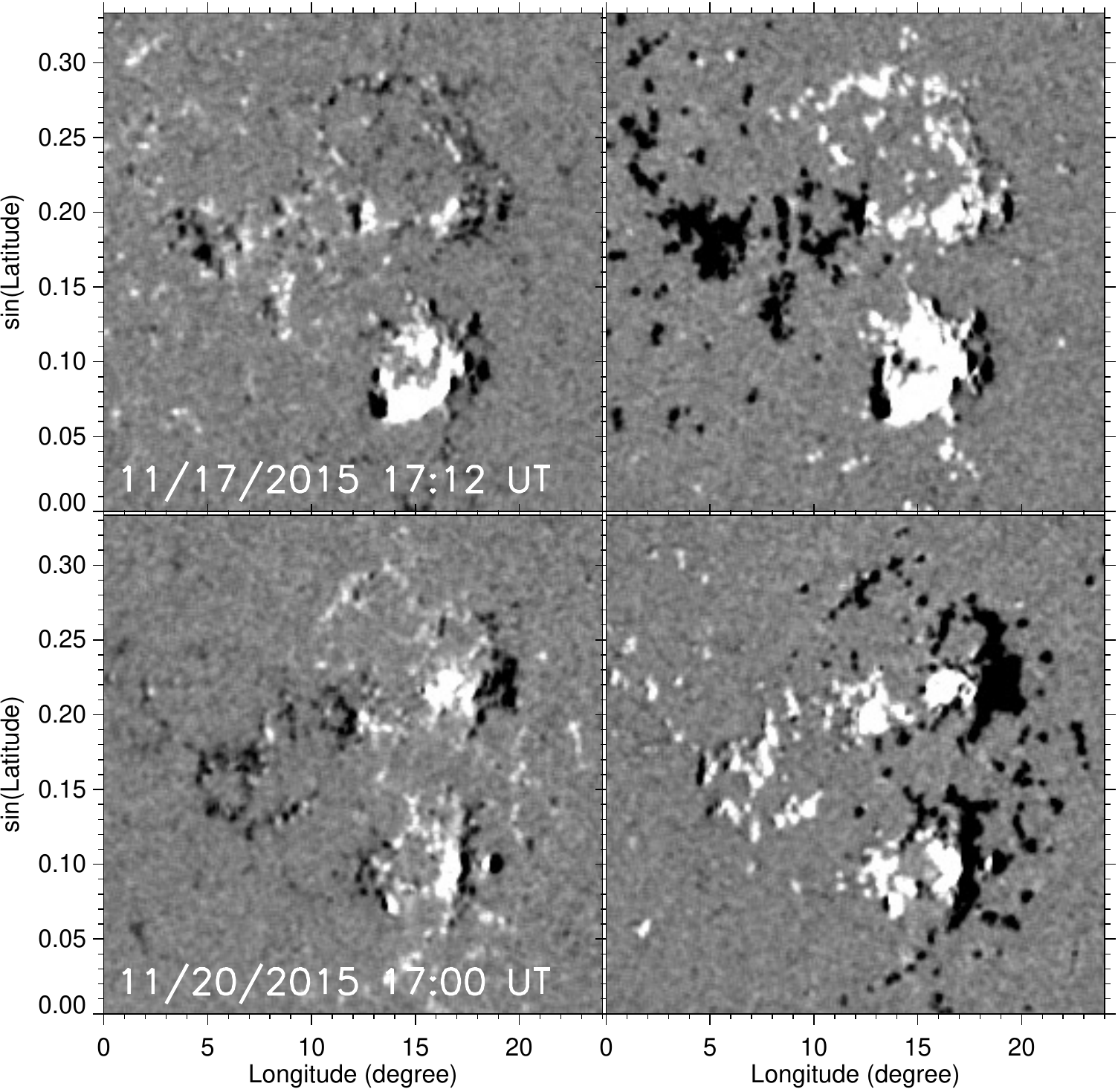}
  \caption{$B_p$ in region A (Figure \ref{fig:fullDisk}) from the new two-component VFISV (left panels) and the original VFISV (right). The data are taken three days apart. The top panels show the data on 17 November when region A is in the east hemisphere; the bottom were observed on 20 November when region A is in the west hemisphere.}
  \label{fig:Aregion}
\end{figure*}

One possible cause for the presence of this hemispheric bias is the fact that a pixel might not be fully occupied by a magnetic atmosphere, i.e. typical magnetic structures outside of sunspots are smaller than the area of an HMI pixel. If this is the case, when performing an inversion assuming that the whole pixel is magnetic (as is the case for current HMI data products that depend on the ME inversion) might lead to an improper magnetic field vector inference, which might be the cause of the sign reverse in $B_p$. This limitation can be alleviated by allowing the inversion to also invert a filling factor that specifies the fraction of the pixel that is to be considered magnetic. It is important to note here that the inclusion of the filling factor, despite being linear, has a different relation to the longitudinal and transverse magnetic field components. For instance, let us take a simple model of the weak field approximation \citep{jefferies1991}. Measurement of the line-of-sight field is proportional to the filling factor ($\alpha_{\rm mag}$), while the transverse field is proportional to $\sqrt{\alpha_{\rm mag}}$. Thus, when $\alpha_{\rm mag}$ is not accurately determined, the various spherical components of the magnetic field vector might reflect spurious effects (such as the hemispheric bias) that are due to the fact that the magnetic field vector is not accurately inferred. It is important to note that in the original VFISV, the filling factor is set to be 1. This is close to the value found in umbrae and penumbrae, but significantly larger than the one found in plages and quiet-Sun regions. Consequently, the effect of the filling factor will mostly affect the inference of the plage - quiet-Sun regions while the penumbrae and umbrae inference will remain unchanged. While we avoid any further discussion on the specifics of the relation between the inference of the magnetic field vector and the presence of the hemispheric bias here, the reader is referred to \cite{liu2021} for a thorough discussion on this topic.

The new version of VFISV with $\alpha_{\rm mag}$ inverted reduces this mismatch, and thus mitigates the hemisphere bias. Figure \ref{fig:Aregion} shows $B_p$ in region A (Figure \ref{fig:fullDisk}) observed three days apart. Left panels are from the two-component VFISV; right are from original VFISV. To avoid projection effects, $B_p$ is mapped onto heliographic coordinates with the X-axis in degrees and Y-axis in sin(Latitude). The region is in the east hemisphere on 17 November (top panels) and west hemisphere on 20 November (bottom panels). The sign change has been mitigated substantially.

\begin{figure*}[ht]
\centering
  \includegraphics[width=0.7\textwidth]{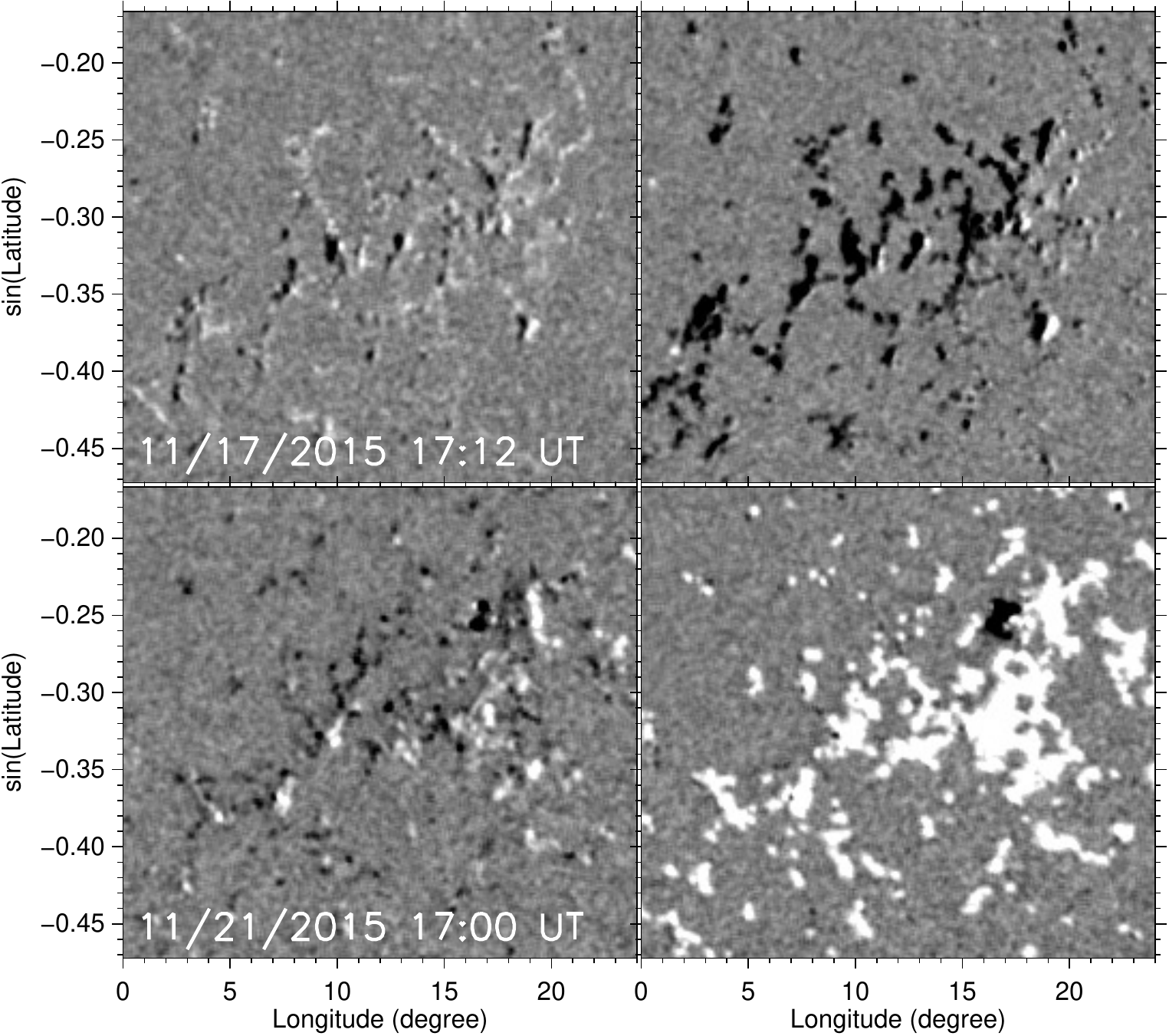}
  \caption{$B_p$ in region B (Figure \ref{fig:fullDisk}) from the two-component VFISV (left panels) and original VFISV (right), taken four days apart, on 17 November (top) when the region is in the east hemisphere, and 21 November (bottom) when the region is in the west hemisphere.}
  \label{fig:Bregion}
\end{figure*}

$B_p$ in region B (Figure \ref{fig:fullDisk}) is shown in Figure \ref{fig:Bregion}. Left panels are from the two-component VFISV and those on the right from the original. Again, the sign change in $B_p$ has been mitigated. More quantitatively, we calculate the percentage of pixels that have $B_p < 0$ for this region. Only the pixels with field strength greater than 300 Mx cm$^{-2}$, $3\sigma$ of the HMI vector data \citep{Hoeksema2014}, are included. The percentage for the original data changes from $(84.8\pm 2.2)\%$ in the east hemisphere (November 17) to $(5.1\pm 1.3)\%$ in the west (November 20). In contrast, it is $(44.3\pm 3.9)\%$ in the east hemisphere and $(43.5\pm 4.0)\%$ in the west for the new data. The error refers to a confidence interval at a 99\% confidence level. The hemisphere bias has been mitigated substantially when the filling factor is inverted with the two-component VFISV.

We can roughly estimate how many pixels in the full-disk vector magnetograms may have a wrong sign in $B_p$. We examine the 17:12 UT 2015 November 17 full disk vector magnetograms in which no active regions are on the disk. About $(22.1\pm 0.2)\%$ pixels have the opposite signs in $B_p$ between the original and new inverted data. The pixels included are over the entire Sun's disk with a confidence index of disambiguation greater than 50. The confidence index denotes confidence level of disambiguation. In high-confidence pixels the minimum energy method is applied; in intermediate confidence pixels the minimum energy result is smoothed using an acute angle method \citep{leka2009,Hoeksema2014}. Pixels with confidence greater than 50 are either high confidence or intermediate. If we assume the new version of VFISV produces true magnetic field, this implies that about 24\% pixels over the disk in the original magnetograms have the wrong sign in $B_p$.

\begin{figure*}[ht]
\centering
  \includegraphics[width=0.65\textwidth]{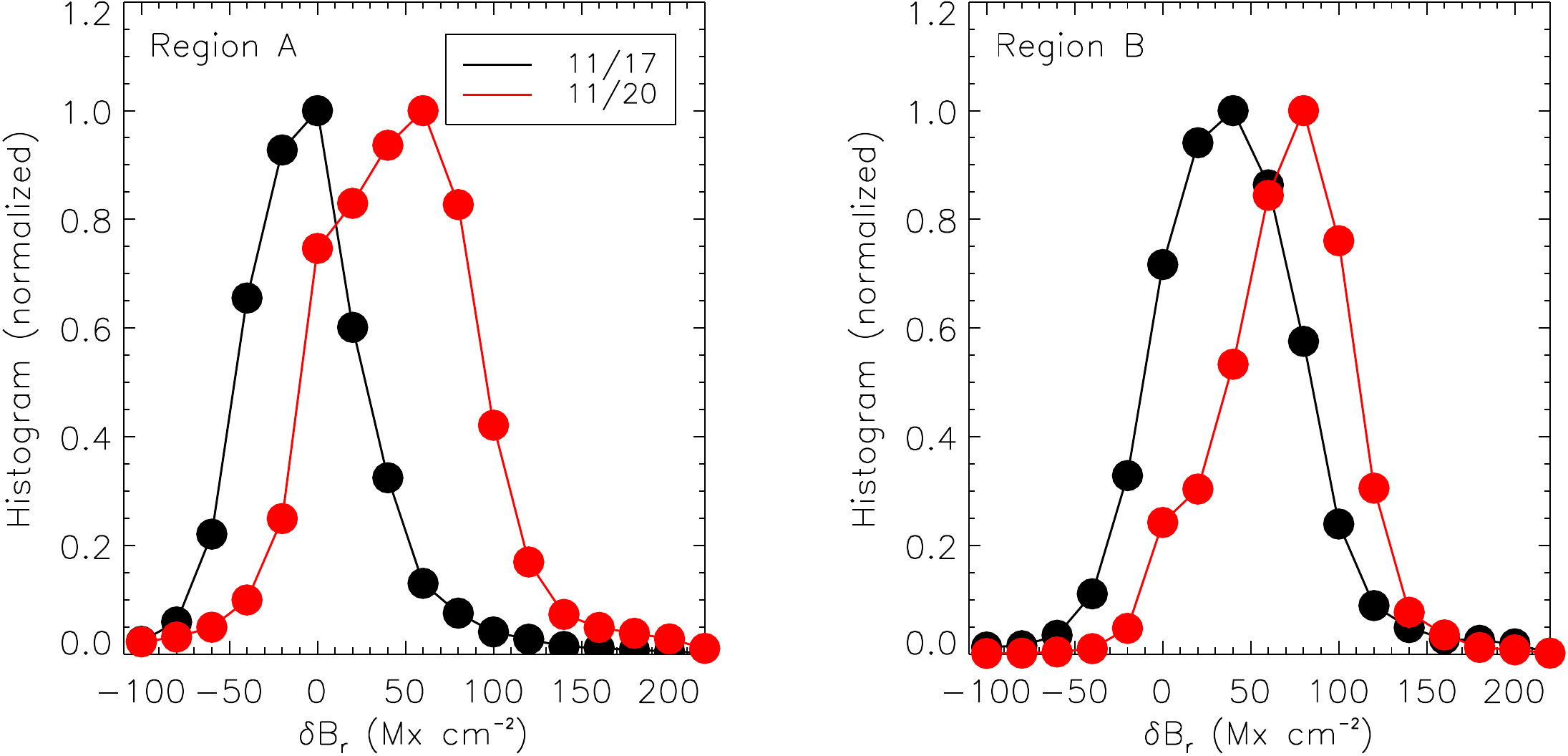}
  \caption{Normalized histograms of $\delta B_r$ for regions A (left panel) and B (right) for the data at 17:12 UT November 17, 2015  (black curves) and 17:00 UT November 20, 2015 (red curves). $\delta B_r = |B_r^{old}| - |B_r^{new}|$, where $B_r^{old}$ refers to radial field from inverted vector magnetic field with original VFISV; $B_r^{new}$ is inverted with the two-component VFISV. Only pixels with $\alpha_{\rm mag} B > 300$ are included.}
  \label{fig:histogram}
\end{figure*}

Vector magnetic field in the spherical coordinates, $B_r$, $B_\theta$, and $B_\phi$, have contributions from observed line-of-sight and transverse fields. $B_r$ may have this hemispheric bias problem, as well. Shown in Figure \ref{fig:histogram} are normalized histograms of $\delta B_r$ for regions A (left panel) and B (right) for the November 17 data (black curves) when the two regions are in the east hemisphere and November 20 data (red curves) when the regions are in the west hemisphere. Here, $\delta B_r = |B_r^{old}| - |B_r^{new}|$. Only the pixels with $\alpha_{\rm mag} B > 300.0$ are included. For both regions, $\delta B_r$ becomes larger when the regions move to the west, implying this hemispheric bias problem also exists in $B_r$ component. Thus correction for this bias is necessary and important. 

\section{Discussion and conclusions} 
\label{sec:conclusions}

We have presented a modification of the original VFISV inversion code to invert the Stokes parameters with a two-component ME model for HMI data in order to better handle intermediate-strong small-scale (unresolved) magnetic structures embedded in non-magnetized (or weakly magnetized) atmosphere. This is motivated because the spatial resolution of the HMI/SDO instrument is not high enough to fully resolve small-scale magnetic structures in the photosphere. 

The two-component model solves for an atmosphere that includes in each pixel both a magnetic component and a non-magnetic component, with the magnetic fraction (filling factor) specified by $\alpha_{\rm mag}$. In order to keep the model complexity as simple as possible, we have chosen to use the same non-magnetic model parameters ($\eta_{0}$, $a$, $\Delta\lambda_{D}$, ${\rm V}_{\rm los}$, $\rm S_{0}$, $\rm S_{1}$) for both the magnetic and non-magnetic contributions. Additionally, the magnetic parameters for the non-magnetic component are set to zero. This way only one additional model parameter is required compared to the standard VFISV model: the filling factor $\alpha_{\rm mag}$.

By means of the Bayesian approach, we have demonstrated that the application of the two-component ME model to weak- and medium-strength magnetic features provides a significantly better inference of the magnetic field vector. Of particular interest are medium field strength features where the linear polarization signals are just strong enough to break the degeneracy of the inferred magnetic field strength, inclination, and filling factor, i.e. outside of strong-field active regions and outside of the quiet-Sun). In this regime, the two-component model is strongly preferable (as deduced from the Bayes factor between both models). The new inversion leads to a different inferred magnetic field vector that is more consistent with the spectropolarimetric observations. We also find that for quiet-Sun pixels with low polarization signals, even if the two-component model is marginally preferable, the additional computational cost together with the fact that the inferred magnetic field vector components are weakly constrained (the actual constraining parameter is the magnetic flux density) discourages the usage of the two-component model. In other words, if this method is to be systematically applied, setting a limit on the total polarimetric signal for which pixels it is useful would inform a smart hybrid systematic inversion of the HMI spectropolarimetric data.

We have shown that the application of this new version of VFISV mitigates the problem of the hemisphere bias in east-west component of magnetic field in regions with intermediate magnetic fields. 
This is because a change in the orientation of the magnetic field vector in the line-of-sight reference frame produces a change in the orientation of the vector in the local one. This change also alters $B_\phi$, improving the value of this component with the strategy shown in this paper.
This hemispheric bias may also affect other components, including $B_r$. For a nominal full-disk vector magnetogram, roughly 22\% of pixels with decent polarization signal might have the wrong sign in $B_p$. The computing time using a lower polarization threshold (0.25 \% in total polarization) increases by about 55\%. Searching for the optimal threshold of polarization and speeding up the code are under investigation.


\acknowledgments

The authors are especially grateful to KD Leka for very interesting discussions. We thank the anonymous referee for valuable comments and suggestions that improve the manuscript. This work was supported by NASA Contract NAS5-02139 (HMI) to Stanford University. This research is also supported by the Research Council of Norway through its Centres of Excellence Scheme, Project Number 262622. This project has received funding from the European Research Council (ERC) under the European Union’s Horizon 2020 research and innovation programme (SUNMAG, grant agreement 759548). The Institute for Solar Physics is supported by a grant for research infrastructures of national importance from the Swedish Research Council (registration number 2017-00625). We acknowledge the community effort devoted to the development of the following open-source packages that were used in this work: 
numpy \citep[numpy.org][]{numpy2020}, 
matplotlib \citep[matplotlib.org][]{hunter2007}. 
The authors are grateful to the SDO/HMI team for their data. 




\bibliography{biblio}{}
\bibliographystyle{aasjournal}


\end{document}